\documentclass{iaus}
\usepackage{amsmath}
\usepackage{amssymb}
\usepackage{mathrsfs}
\usepackage{graphicx}

\newcommand{\rmd}{{\mathrm{d}}}
\newcommand{\Msun}{\:{\rm M}_{\odot}}

\newcommand{\Mtwohun}{{\rm M}_{\rm halo}}
\newcommand{\rhalf}{{\rm r}_{_{1/2}}}
\newcommand{\Rhalf}{{\rm R}_{_{\rm e}}}

\newcommand{\rlimit}{{\rm r}_{_{\rm lim}}}
\newcommand{\rcut}{{\rm r}_{_{\rm cut}}}
\newcommand{\rmax}{{\rm r}_{_{\rm high}}}
\newcommand{\rbeta}{{\rm r}_{_{\beta}}}

\newcommand{\req}{{\rm r}_{_{\rm eq}}}

\newcommand{\Mhalf}{{\rm M}_{_{1/2}}}
\newcommand{\Lhalf}{{\rm L}_{_{1/2}}}
\newcommand{\moverl}{\Upsilon^{{\rm I}}_{_{1/2}}}
\newcommand{\moverlI}{\Upsilon^{{\rm I}}_{_{1/2}}}
\newcommand{\sigmalos}{\sigma_{_{\rm{los}}}}

\newcommand{\Lsun}{{\rm L}_\odot}

\newcommand{\LV}{{\rm L}_{\rm V}}

\newcommand{\LI}{{\rm L}_{\rm I}}
\newcommand{\kms}{{\rm km \: s}^{-1}}
\newcommand{\LCDM}{\Lambda{\rm CDM}}
\newcommand{\beq}{\begin{equation}}
\newcommand{\eeq}{\end{equation}}

\newcommand{\data}{{\mathscr{V}}}
\newcommand{\dataD}{{\mathscr{D}}}
\newcommand{\Einastop}{{\mathscr{M}}}

\newcommand{\avelos}{\langle \sigma_{_{\rm los}}^2 \rangle}

\title[Modeling mass independent of anisotropy] 
{Modeling mass independent of anisotropy}

\author[Joe Wolf]   
{Joe Wolf}

\affiliation{Center for Cosmology, Department of Physics \& Astronomy \\ University of California, Irvine, CA 92697 \\ wolfj@uci.edu
}

\pubyear{2010}
\volume{271}  
\pagerange{001--009}
\jname{Astrophysical Dynamics: From Stars to Galaxies}
\editors{N. Brummell \& A.S. Brun eds}
\begin{document}

\maketitle

\begin{abstract}
By manipulating the spherical
Jeans equation, \cite{Wolf10} show that the mass enclosed within the 
3D deprojected half-light radius $\rhalf$ can be determined with only mild 
assumptions about the spatial variation of the stellar velocity dispersion anisotropy
as long as the projected velocity dispersion profile is fairly flat near the 
half-light radius, as is typically observed. They find $\Mhalf = 3 \, G^{-1} \, 
\avelos \, \rhalf \simeq 4 \, G^{-1} \, \avelos \, \Rhalf$, where
$\avelos$ is the luminosity-weighted square of the 
line-of-sight velocity dispersion and $\Rhalf$ is the 2D projected
half-light radius. This finding can be used to show that all of the Milky Way dwarf
spheroidal galaxies (MW dSphs) are consistent with having formed 
within a halo of mass approximately $3 \times 10^9 \Msun$, assuming a
$\LCDM$ cosmology. In addition, the dynamical I-band
mass-to-light ratio $\moverlI$ vs. $\Mhalf$ relation for dispersion-supported galaxies
follows a U-shape, with a broad minimum near $\moverlI \simeq 3$ that
spans dwarf elliptical galaxies to normal ellipticals, a steep
rise to $\moverlI \simeq$ 3,200 for ultra-faint dSphs, and a more
shallow rise to $\moverlI \simeq 800$ for galaxy cluster spheroids.  
\end{abstract}

\firstsection 
\section{Introduction}

Mass determinations for dispersion-supported galaxies based on only line-of-sight 
velocity measurements suffer from an uncertainty associated with not 
knowing the intrinsic 3D velocity dispersion. The difference between tangential and radial 
velocity dispersions is quantified by the stellar velocity dispersion anisotropy, 
$\beta$. Many questions in galaxy formation are affected by our ignorance of $\beta$, 
including the ability to quantify the dark matter content in the outer parts of elliptical 
galaxies [\cite{Romanowsky_03, Dekel_05}], to measure the mass profile of the Milky 
Way from stellar halo kinematics [\cite{Battaglia_05, Dehnen_06}], and to infer accurate 
mass distributions in dwarf spheroidal galaxies (dSphs) [\cite{Gilmore_07, Strigari_07}].

\cite{Wolf10} (hereafter W10) used the spherical Jeans equation to show that for each
dispersion-supported galaxy, there exists one radius within which the
integrated mass as inferred from the line-of-sight velocity dispersion
is largely insensitive to $\beta$, and that for a wide range of stellar
distributions, this radius is approximately the 3D deprojected half-light 
radius $\rhalf$:
\begin{eqnarray}
\label{eq:main}
\Mhalf & \equiv& M(\rhalf) \simeq 3 \, G^{-1} \, \avelos\, \rhalf \,, \\
& \simeq & 4 \, G^{-1} \, \avelos \, \Rhalf \,,\nonumber\\
& \simeq & 930 \: \left(\frac{\avelos}{\rm km^2 \, s^{-2}} \right)
 \: \left(\frac{\Rhalf}{\rm pc}\right) \: \Msun\,. \nonumber 
\end{eqnarray}
$\avelos$ is the luminosity-weighted square of the 
line-of-sight velocity dispersion.
In the second line we have used $\Rhalf \simeq (3/4) \,
\rhalf$ for the 2D projected half-light radius. This approximation is
accurate to better than 2\% for exponential, Gaussian, King, Plummer,
and S\'ersic profiles (see Appendix B of W10 for useful fitting formulae).

\section{The Spherical Jeans Equation}
\label{sec:JE}

Given the relative weakness of the scalar virial theorem as a mass estimator (see Section 2 of W10), we turn to the spherical Jeans equation.
It relates the tracer velocity dispersion and tracer number density $n_{\star}(r)$ of a spherically symmetric, dispersion-supported, collisionless stationary system to its total gravitating potential $\Phi(r)$, under the assumption of dynamical equilibrium with no streaming motions:
\begin{equation}
\label{eq:jeans}
- n_{\star} \frac{\rmd \Phi}{\rmd r} = \frac{\rmd(n_{\star} \sigma_r^2)}{\rmd r} + 2 \frac{\beta \: n_{\star} \sigma_r^2}{r}.
\end{equation}
$\sigma_r(r)$ is the radial velocity dispersion of the stars/tracers and $\beta(r) \equiv 1- \sigma_t^2 / \sigma_r^2$ measures the velocity anisotropy, where the tangential velocity dispersion $\sigma_t =\sigma_\theta = \sigma_\phi$. It is informative to rewrite the implied total mass profile as
\begin{equation}
\label{eq:massjeans}
M(r) = \frac{r \: \sigma_r^2}{G} \left (\gamma_\star+\gamma_\sigma - 2\beta \right ),
\end{equation}
where $\gamma_{\star} \equiv - \rmd \ln n_\star / \rmd \ln r$ and $\gamma_{\sigma} \equiv - \rmd \ln \sigma_r^2 / \rmd \ln r$. 
Given line-of-sight kinematics, the only term on the right-hand side of Equation \ref{eq:massjeans}
that can be determined by observations is $\gamma_\star$, which follows from the projected surface brightness 
profile under an assumption about how it is related to the projected stellar number density $\Sigma_\star(R)$. 
Via an Abel inversion $n_\star$ is mapped in a one-to-one manner with the spherically deprojected observed surface 
brightness profile by assuming that $n_\star$ traces the light density.

Line-of-sight kinematic data provides the projected velocity dispersion profile $\sigmalos(R)$. 
As first shown by \cite{BinneyMamon_82}, in order to use the Jeans equation one must relate
 $\sigmalos$ to $\sigma_r$:
\begin{equation}
\label{eq:LOSrelation}
\Sigma_\star \, \sigmalos^2(R)  =  \int^{\infty}_{R^2} n_\star \sigma_r^2(r) \left[1 - \frac{R^2}{r^2}\beta(r)\right] \frac{\rmd r^2}{\sqrt{r^2 - R^2}}.
\end{equation}
It is clear then that a significant degeneracy associated with using the observed $\Sigma_\star(R)$ and $\sigmalos(R)$ profiles exists in trying to determine an underlying mass profile $M(r)$ at any radius, as uncertainties in $\beta$ will affect both the relationship between $M(r)$ and $\sigma_r$ in Equation \ref{eq:massjeans} and the mapping between $\sigma_r$ and $\sigmalos$ in Equation \ref{eq:LOSrelation}.

The technique of W10 for handling the $\beta$ degeneracy in order to provide a fair representation of the allowed mass profile given a set of observables is to consider general parameterizations for $M(r)$ and $\beta(r)$ and then to undertake a maximum likelihood analysis to constrain all possible parameter combinations. By using such a strategy, W10 derive meaningful mass likelihoods for a number of dispersion-supported galaxies with line-of-sight velocity data sets.

The stellar velocity dispersion anisotropy can be modeled as a three-parameter function
\begin{equation}
\beta (r) = (\beta_\infty - \beta_0) \frac{r^2}{r^2 + \rbeta^2} + \beta_0, 
\label{eq:betaprofile}
\end{equation}
and the total mass density distribution can be described using the six-parameter function
\begin{equation} 
\rho_{\rm tot}(r) = \frac{\rho_s \, e^{-r/\rcut}}{(r/r_s)^\gamma [1+(r/r_s)^\alpha]^{(\delta-\gamma)/\alpha}}. 
\label{eq:rhor}
\end{equation}
For their marginalization, W10 adopt uniform priors over the following
ranges: $\log_{10}(\rhalf / 5) < \log_{10}(\rbeta) < \log_{10}(\rlimit)$;
$-10 < \beta_\infty < 0.91$; $-10 < \beta_0 < 0.91$;
$\log_{10}(\rhalf / 5) < \log_{10}(r_s) < \log_{10}(2 \, \rmax)$;
$0 < \gamma < 2$; $3 < \delta < 5$; and $0.5 < \alpha < 3$, where
$\rlimit$ is the truncation radius for the stellar density.
The variable $\rcut$ allows the dark matter halo profile to truncate beyond the stellar extent  
and the uniform prior $\log_{10}(\rlimit) < \log_{10}(\rcut) < \log_{10}(\rmax)$
is adopted. For distant galaxies W10 use $\rmax = 10 \, \rlimit$ and for satellite 
galaxies of the Milky Way they set $\rmax$ equal to the Roche limit for a 
$10^9 \Msun$ point mass. In practice, this allowance for $\rcut$ is not 
important when focusing on integrated masses within the stellar radius.
  
%
\begin{figure*}
\includegraphics[width=64mm]{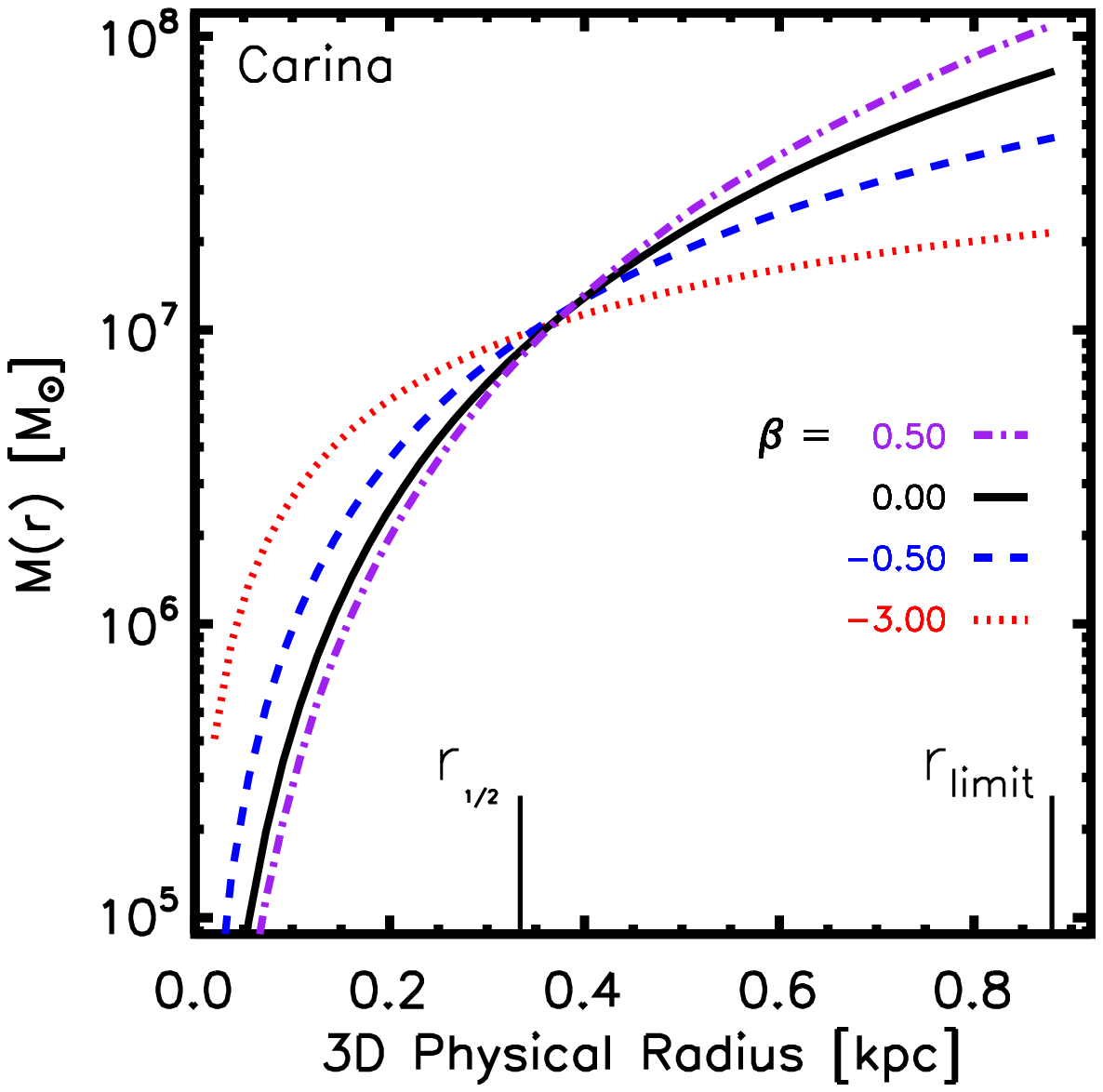}
\hspace{3mm}
\includegraphics[width=64mm]{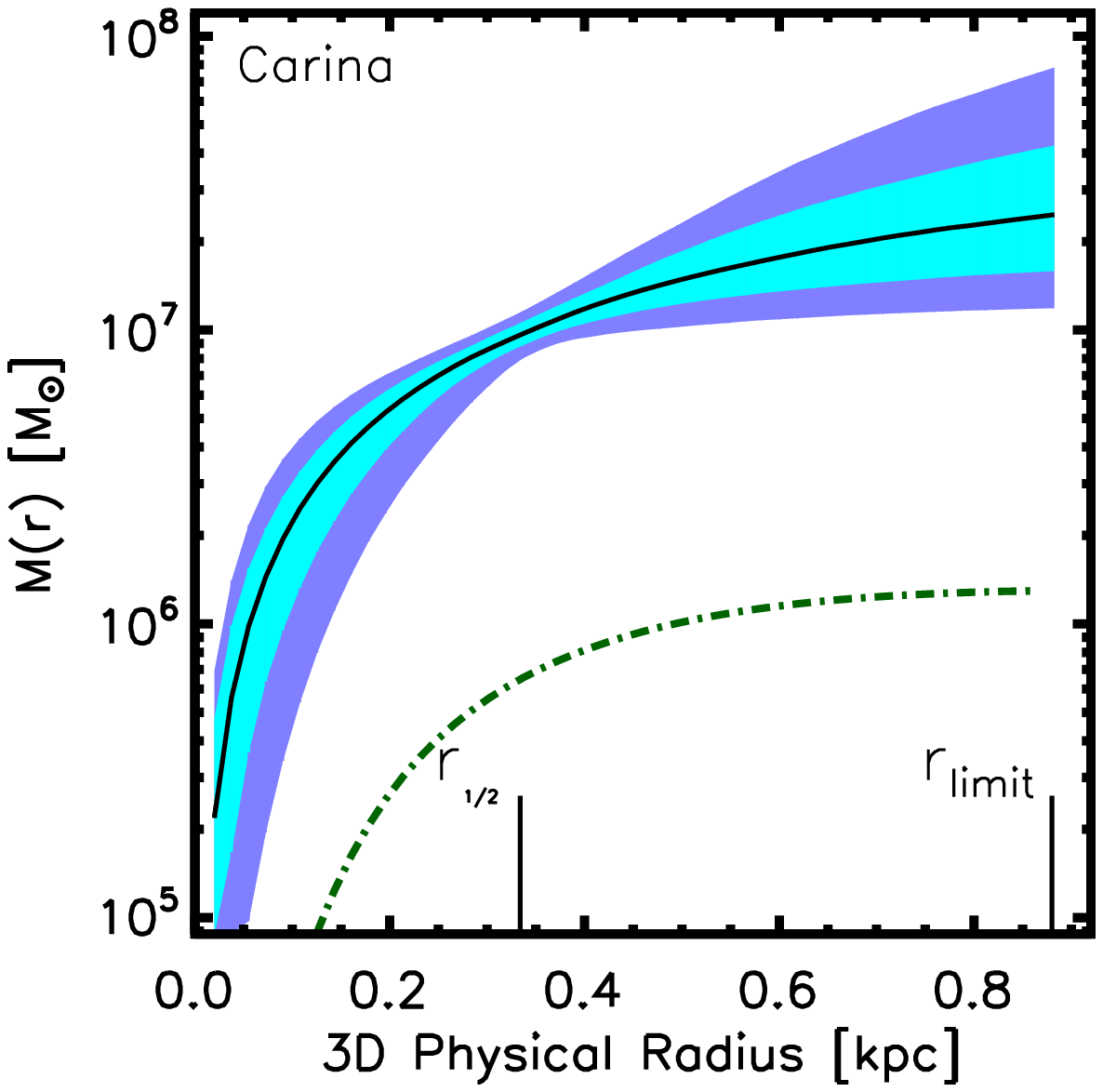}
\caption{Figure 1 from W10. See the text for details. {\em Right:} The green dot-dashed line represents the contribution of mass from the stars, assuming a stellar V-band mass-to-light ratio of 3 $\Msun / \Lsun$.
}
\label{fig:multibeta}
\end{figure*} 
%
  
W10 apply their marginalization procedure to resolved kinematic 
data for MW dSphs, MW globular clusters, and elliptical galaxies. 
Since MW dSphs and globular clusters are close enough for individual
stars to be resolved, the joint probability of obtaining
each observed stellar velocity given its observational error and the
predicted line-of-sight velocity dispersion from Equations
\ref{eq:jeans} and \ref{eq:LOSrelation} is considered.
In modeling the line-of-sight velocity distribution for any system, 
the observed distribution is a convolution
of the intrinsic velocity distribution, arising from the distribution
function, and the measurement uncertainty from each individual star.  
Given that line-of-sight velocity distributions of dispersion-supported
systems are often well-described by a Gaussian, the probability 
of obtaining a set of line-of-sight velocities $\data$ given a set 
of model parameters ${\Einastop}$ is described by the likelihood 

\begin{equation}
\label{eq:fulllike}
P(\data| \Einastop) =  \prod_{i=1}^{N} 
\frac{1}{\sqrt{2\pi(\sigma_{\rm th, i}^2 + \epsilon_i^2)}} 
\exp \left[-\frac{1}{2}\frac{(\data_i - \bar{v})^2}{\sigma_{\rm th, i}^2 
+ \epsilon_i^2}\right]\,. 
\end{equation}

The product is over the set of $N$ stars, where $\bar{v}$ is the average velocity of the galaxy. As expected, the total error at a projected position is a sum in quadrature of the theoretical intrinsic dispersion, $\sigma_{\rm th, i}(\Einastop)$, and the measurement error $\epsilon_i$. The posterior probability distribution for the mass at any radius can be generated by multiplying the likelihood by the prior distribution for each of the nine $\rho_{\rm tot}(r)$ and $\beta(r)$ parameters as well as the observationally derived parameters and associated errors that yield $n_\star(r)$ for each galaxy, including uncertainties in distance. We then integrate over all model parameters, including $\bar{v}$, to derive a likelihood for mass. Following \cite{Martinez_09}, a Markov Chain Monte Carlo technique is used in order to perform the required ten to twelve dimensional integral.

For elliptical galaxies that are located too far for individual stellar spectra to be obtained, the resolved dispersion profiles are analyzed with the likelihood
\begin{equation}
\label{eq:dispersionlike}
P(\dataD| \Einastop) = \prod_{i=1}^{N}
\frac{1}{\sqrt{2\pi}\epsilon_i}
\exp \left[-\frac{1}{2}\frac{(\dataD_i - \sigma_{th, i})^2}{\epsilon_i^2}\right]\, ,
\end{equation}
where the product is over the set of $N$ dispersion measurements $\dataD$, and $\epsilon_i$ is the reported error of each measurement. 

\section{Minimizing the Anisotropy Degeneracy}
\label{sec:mass}

As discussed in W10, the degeneracy between 
the anisotropy and the integrated mass will be 
minimized at an intermediate radius within the stellar distribution. 
Such an expectation follows from considering the relationship 
between $\sigma_r$ and $\sigmalos$.

At the observed center of a spherical, dispersion-supported galaxy
($R = 0$), line-of-sight velocities project onto the radial
component with $\sigmalos \sim \sigma_r$, while at the edge of the
system ($R = \rlimit$), line-of-sight velocities project onto the
tangential component with $\sigmalos \sim \sigma_t$. As an example,
consider an intrinsically isotropic galaxy ($\beta = 0$). If
this system is analyzed using line-of-sight kinematics under the false
assumption that $\sigma_r < \sigma_t$ ($\beta < 0$) at all radii, then 
the total velocity dispersion at $r \simeq 0$ would be overestimated 
while the total velocity dispersion at $r \simeq \rlimit$ would be
underestimated. Conversely, if one were to analyze the same galaxy
under the assumption that $\sigma_r > \sigma_t$ ($\beta > 0$) at all radii,
then the total velocity dispersion would be underestimated near the center
and overestimated near the galaxy edge. Thus, 
some intermediate radius should exist where attempting to infer the
enclosed mass from only line-of-sight velocities is minimally affected
by the unknown value of $\beta$.

These qualitative expectations are quantitatively displayed in Figure
\ref{fig:multibeta}, where inferred mass profiles for the Carina dSph 
galaxy for several choices of constant $\beta$ are shown. The
right-hand panel shows the same data analyzed using the complete
likelihood analysis, where the fairly general $\beta(r)$ profile presented
in Equation \ref{eq:betaprofile} was marginalized over.
The dataset is discussed in W10, where the average velocity error is
$\sim 3 \, \kms$. Each line in the left panel of Figure \ref{fig:multibeta} shows
the median likelihood of the cumulative mass value at each radius for
the value of $\beta$ indicated. The 3D half-light radius and the limiting
stellar radius are marked for reference. As expected, forcing $\beta < 0$
produces a systematically higher (lower) mass at a small (large)
radius compared to $\beta > 0$. Thus, this requires that every pair
of $M(r)$ profiles analyzed with different assumptions about $\beta$ cross 
at some intermediate radius.
Somewhat remarkable is that there exists a radius where every pair 
approximately intersects. We see that this radius is very close to 
the deprojected 3D half-light radius $\rhalf$. The right-hand panel in Figure
\ref{fig:multibeta} shows the full mass likelihood as a function of
radius, with the shaded bands illustrating the 68\% and 95\% likelihood contours.
The likelihood contour also pinches near
$\rhalf$, as the data preferentially constrain this mass value. 

By examining each of the well-sampled dSph kinematic data sets
in more detail, W10 finds that the measurement errors, rather than the 
$\beta$ uncertainty, always dominate the errors on the mass near $\rhalf$,
while the mass errors at {\em both} smaller and larger radii are
dominated by the $\beta$ uncertainty (and thus are less affected by
measurement error). To analytically describe this effect, let us briefly 
examine the Jeans equation in the context of observables. 

Consider a dispersion-supported stellar system where
$\Sigma_\star(R)$ and $\sigmalos(R)$ are determined accurately by
observations, such that any viable solution will keep the quantity 
$\Sigma_\star(R) \, \sigmalos^2(R)$ fixed to within allowable errors. 
With this in mind, W10 show that Equation \ref{eq:LOSrelation} can 
be rewritten in a form that is invertible, isolating into a kernel the
integral's $R$-dependence: 

\begin{eqnarray}
\label{eq:LOSstep}
\Sigma_\star \sigmalos^2 (R) & = & \int^{\infty}_{R^2} n_\star \sigma_r^2(r) \left[1 - \frac{R^2}{r^2}\beta(r)\right] \frac{\rmd r^2}{\sqrt{r^2 - R^2}} \hspace{.4in}
\\ \nonumber
& = & \int_{R^2}^{\infty} \frac{n_\star \sigma_r^2}{r^2}\frac{(1 - \beta)r^2 + \beta(r^2 - R^2)}{\sqrt{r^2-R^2}} \rmd r^2 \\ \nonumber
& = & \int_{R^2}^{\infty} \frac{n_\star \sigma_r^2(1 - \beta)}{\sqrt{r^2-R^2}} \rmd r^2
- \left.\left(\sqrt{r^2-R^2} \int_{r^2}^\infty \frac{\beta n_\star \sigma_r^2}{\tilde{r}^2} \rmd \tilde{r}^2\right)\right|^\infty_{R^2} \\ \nonumber
& + & \int_{R^2}^\infty\left(\int_{r^2}^\infty \frac{\beta n_\star \sigma_r^2}{\tilde{r}^2} \rmd \tilde{r}^2\right)\frac{1}{2}\frac{\rmd r^2}{\sqrt{r^2-R^2}} \\ \nonumber
& = & \int_{R^2}^{\infty} \left[\frac{n_\star \sigma_r^2}{(1 - \beta)^{-1}} + \int_{r^2}^\infty \frac{\beta n_\star \sigma_r^2}{2\tilde{r}^2} \rmd \tilde{r}^2 \right] \frac{\rmd r^2}{\sqrt{r^2-R^2}} \, , 
\end{eqnarray}
where an integration by parts was utilized to achieve the third equality. Note that the 
second term on the third line must be zero under the physically-motivated assumption that
the combination $\beta n_\star \sigma_r^2$ falls faster than $r^{-1}$ at large r.
\footnote{Appendix A of W10 shows how an Abel inversion can be used to solve for 
$\sigma_r(r)$ and $M(r)$ in terms of directly observable quantities.}

{\bf Because Equation \ref{eq:LOSstep} is invertible, the fact that the
left-hand side is an observed quantity and independent of $\beta$
implies that the term in brackets must be well determined regardless 
of a chosen $\beta$.} This allows one to equate the isotropic integrand
with an arbitrary anisotropic integrand: 
\beq
\label{eq:integrand}
 \left . n_\star \sigma_r^2 \right |_{\beta = 0} = n_\star
 \sigma_r^2 [1-\beta(r)] + \int_r^\infty \frac{\beta n_\star
   \sigma_r^2 \rmd \tilde{r}}{\tilde{r}}. 
\eeq
After taking a derivative with respect to $\ln r$ and subtracting Equation 
\ref{eq:jeans} the following result is obtained
\begin{eqnarray}
\label{eq:massdiff}
M(r; \beta) - M(r; 0) = \frac{\beta(r) \: r \: \sigma_r^2(r)}{G} 
\left( \gamma_\star + \gamma_\sigma + \gamma_\beta - 3 \right).
\end{eqnarray}
We remind the reader that $\gamma_\star \equiv - \rmd \ln n_\star / \rmd \ln r$ 
and $\gamma_\sigma \equiv - \rmd \ln \sigma_r^2 / \rmd \ln r$. Likewise,
 $\gamma_\beta \equiv - \rmd \ln \beta / \rmd \ln r = - \beta^\prime/\beta$, 
where $^\prime$ denotes a derivative with respect to $\ln r$.
The possibility of a radius $\req$ is revealed by Equation \ref{eq:massdiff},
where the term in parentheses goes to zero, such that the enclosed mass 
$M(\req)$ is minimally affected by not knowing 
$\beta(r)$:
\begin{equation}
\label{eq:req}
\gamma_\star(\req) = 3 - \gamma_\sigma(\req) - \gamma_\beta(\req) \,.
\end{equation}

W10 go through additional detailed arguments to justify that $\req \simeq \rhalf$ for systems
with relatively flat observed dispersion profiles. We will utilize this finding (Equation 
\ref{eq:main}) to perform tests of galaxy formation.

%
\begin{figure*}
\includegraphics[width=64mm]{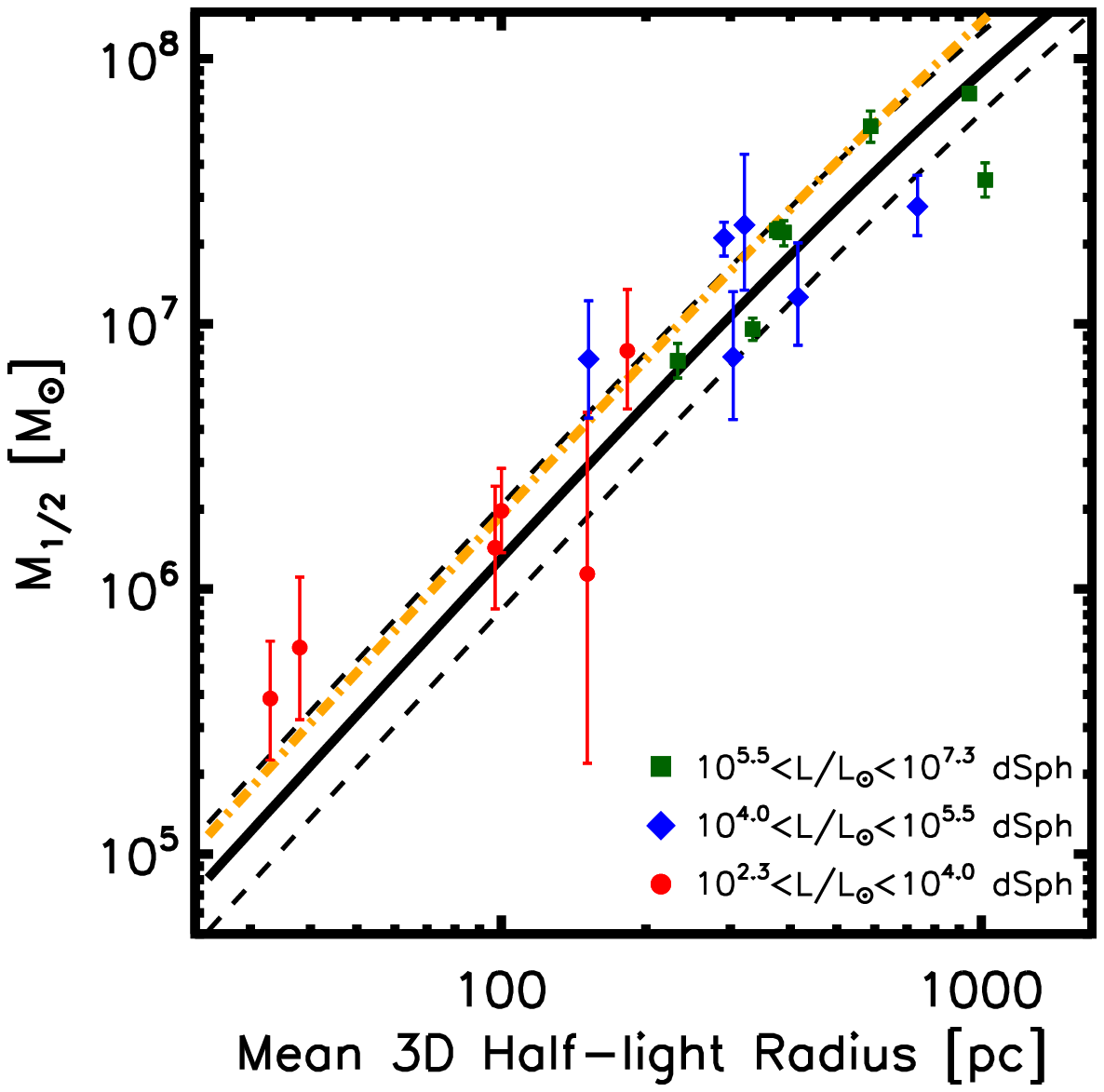}
\hspace{3mm}
\includegraphics[width=64mm]{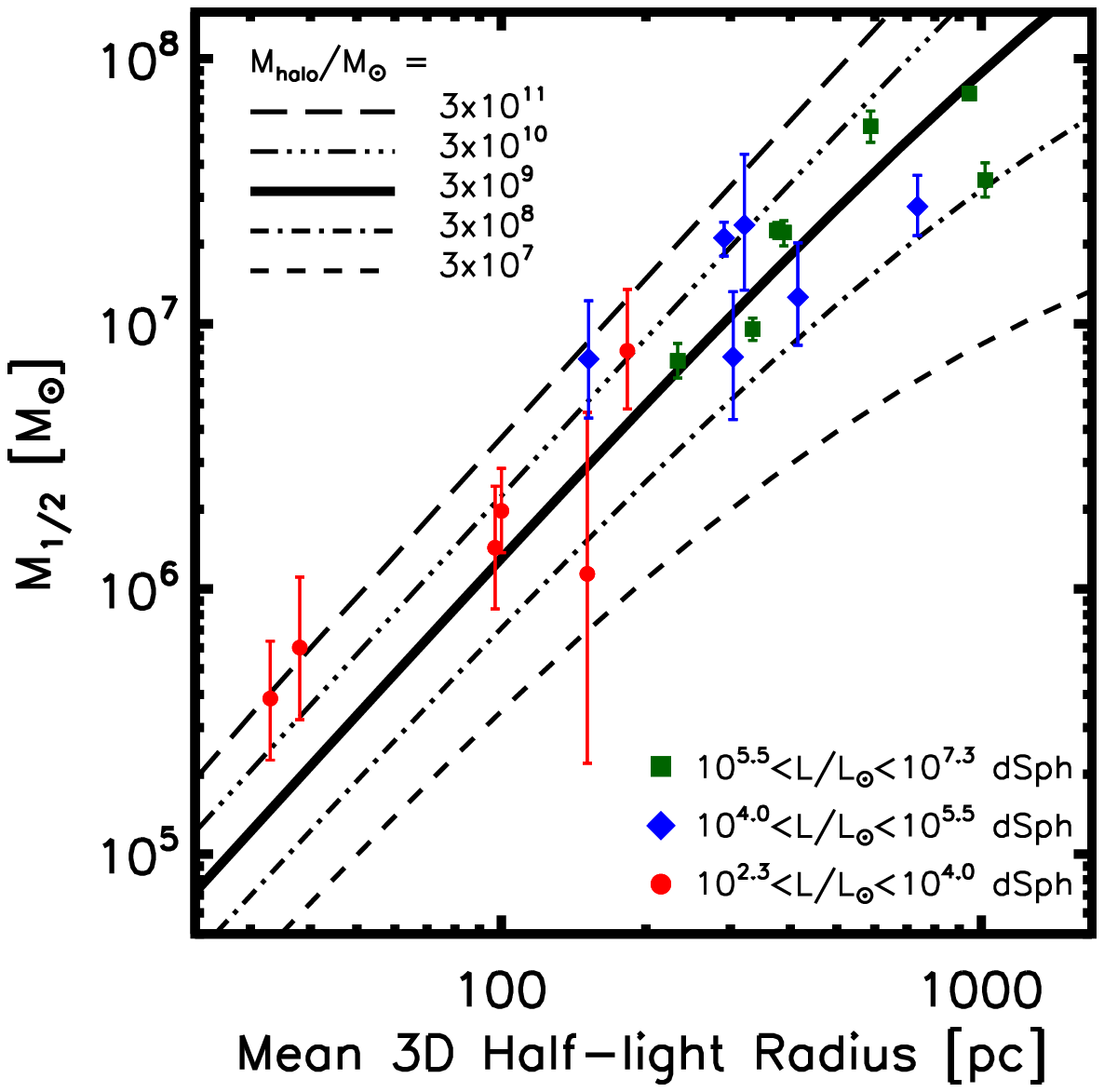}
\caption{Figure 3 from W10. The half-light masses of the Milky Way dSphs plotted against $\rhalf$. See the text for descriptions.}
\label{fig:Mrhalfvsrhalf}
\end{figure*} 
%

\section{Milky Way dwarf spheroidal galaxies}

As an example of the utility of $\Mhalf$ determinations, 
Figure \ref{fig:Mrhalfvsrhalf} presents $\Mhalf$ vs. $\rhalf$ for MW
dSph galaxies. Relevant parameters for each of the galaxies
are provided in Table 1 of W10. The symbol types labeled on the plot
correspond to three luminosity bins that span almost five orders
of magnitude in luminosity. It is interesting to compare the data points in Figure
\ref{fig:Mrhalfvsrhalf} to the integrated mass profile $M(r)$
predicted for $\Lambda$CDM dark matter field halos of a given $\Mtwohun$
mass, which is defined as the halo mass corresponding to an
overdensity of 200 compared to the critical density. In
the limit that dark matter halo mass profiles $M(r)$ map 
in a one-to-one way with their $\Mtwohun$ mass, then the
points on this figure may be used to estimate an associated halo mass
for each galaxy. 

The solid line in the left panel of Figure \ref{fig:Mrhalfvsrhalf}
shows the mass profile for a NFW dark matter halo [\cite{nfw}] at
$z=0$ with a halo mass  $\Mtwohun = 3 \times 10^9 \Msun$. 
The median concentration ($c=11$) predicted by the mass-concentration model 
[\cite{Bullock_01}] updated for WMAP5 $\Lambda$CDM parameters [\cite{Maccio_08}]
is used. The dashed lines
indicate the expected $68 \%$ scatter about the median concentration
at this mass. The dot-dashed line shows the expected $M(r)$ profile
for the same mass halo at $z=3$ (corresponding to a concentration of $c=4$), 
which provides an idea of the spread that would result from the 
scatter in infall times. An interesting result is that each MW dSph 
is consistent with inhabiting a dark matter halo of mass 
$\sim 3 \times 10^9 \Msun$ [\cite{Strigari_08}].

The right panel in Figure \ref{fig:Mrhalfvsrhalf} shows the same data
plotted with the median mass profiles for different halo
masses. Clearly, the MW dSphs are also consistent with populating
dark matter halos of a wide range in $\Mtwohun$ above $\sim 3 \times 10^8 \Msun$.
This result informs of a very stringent limit on the fraction of the baryons 
converted to stars in these
systems. More importantly, there is no systematic correlation
between luminosity and the $\Mtwohun$ mass profile that each dSph
most closely intersects\footnote{There are hints by \cite{Kalirai_10} that 
this result does not hold for the M31 dSphs.}. The ultra-faint dSph 
population (circles) with $\LV < 10^4 \, \Lsun$ is equally likely to be 
associated with the more massive dark matter halos as are classical 
dSphs that are more than three orders of magnitude brighter (squares).
A simple interpretation of the right-hand panel of Figure
\ref{fig:Mrhalfvsrhalf} shows that the two least luminous satellites
(which also have the smallest $\Mhalf$ and $\rhalf$ values) are
associated with halos that are {\em more massive} than any of the
classical MW dSphs\footnote{\cite{Martinez_10}, \cite{Minor_10}, 
and \cite{Simon_10} have explored the effects of the inflation of
the observed velocity dispersion due to binaries, but find that
the faintest system, Segue 1, is still extremely dark-matter dominated.}.
This behavior is difficult
to explain in models constructed to reproduce the Milky Way satellite
population, which typically predict a trend between dSph luminosity
and halo infall mass. It is possible that a new scale in galaxy formation 
exists [\cite{Strigari_08}] or that there is a systematic bias that makes 
it difficult to detect low luminosity galaxies which sit within low mass
halos [\cite{Bullock_10}].
\section{Global population of dispersion-supported systems}
\begin{figure*}
\includegraphics[width=44mm]{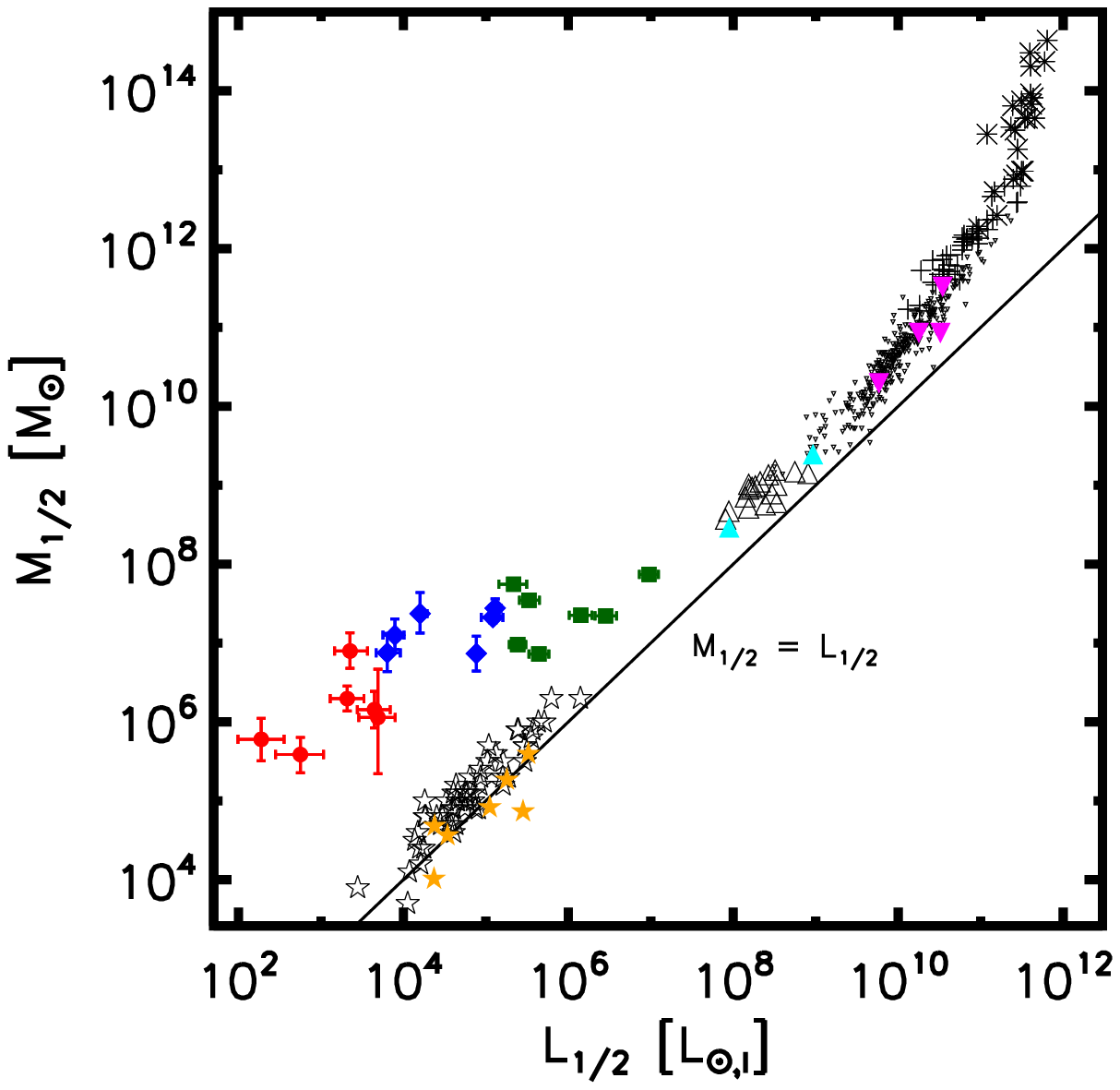}
\hspace{.5mm}
\includegraphics[width=43mm]{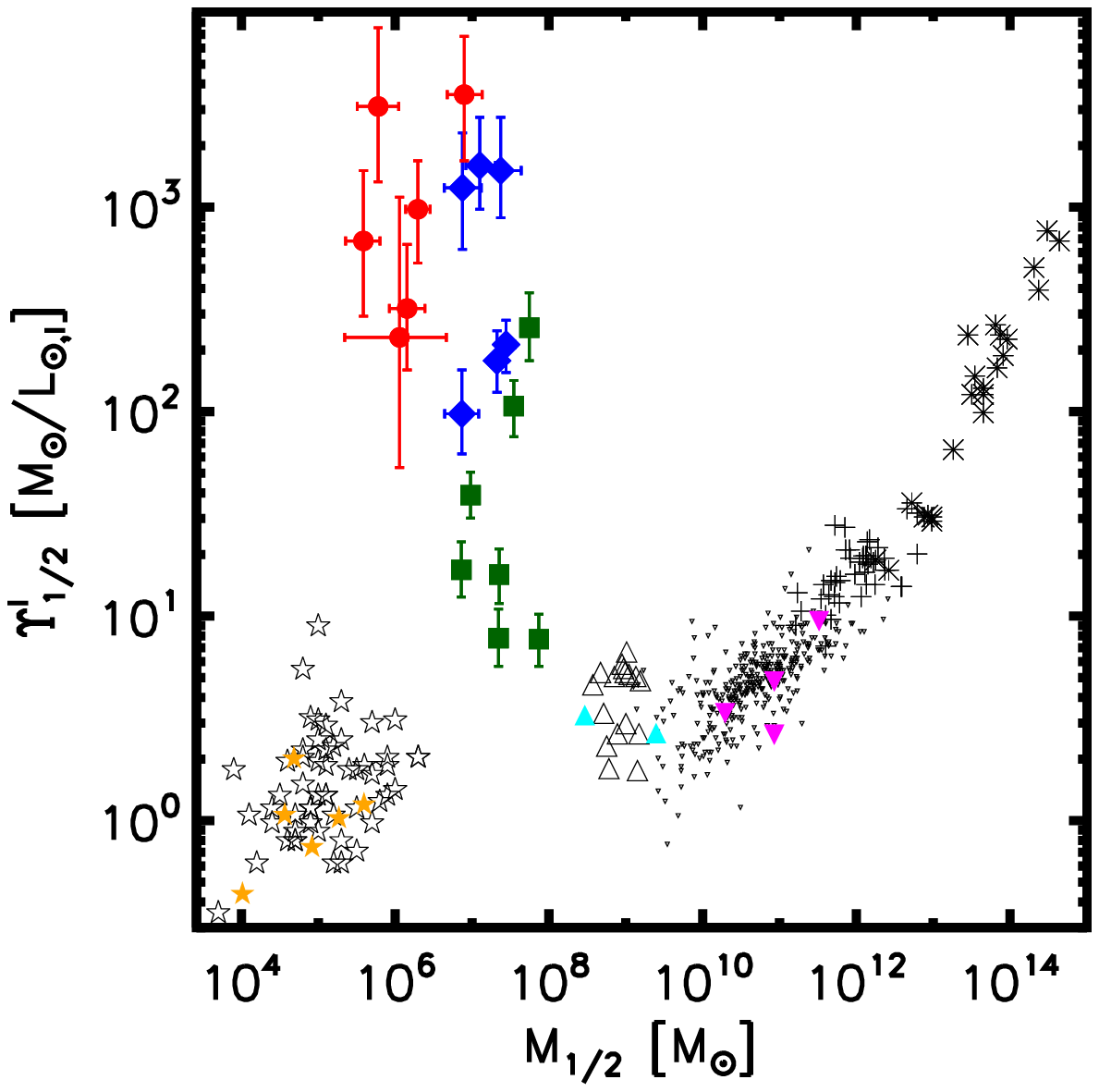}
\hspace{.5mm}
\includegraphics[width=43mm]{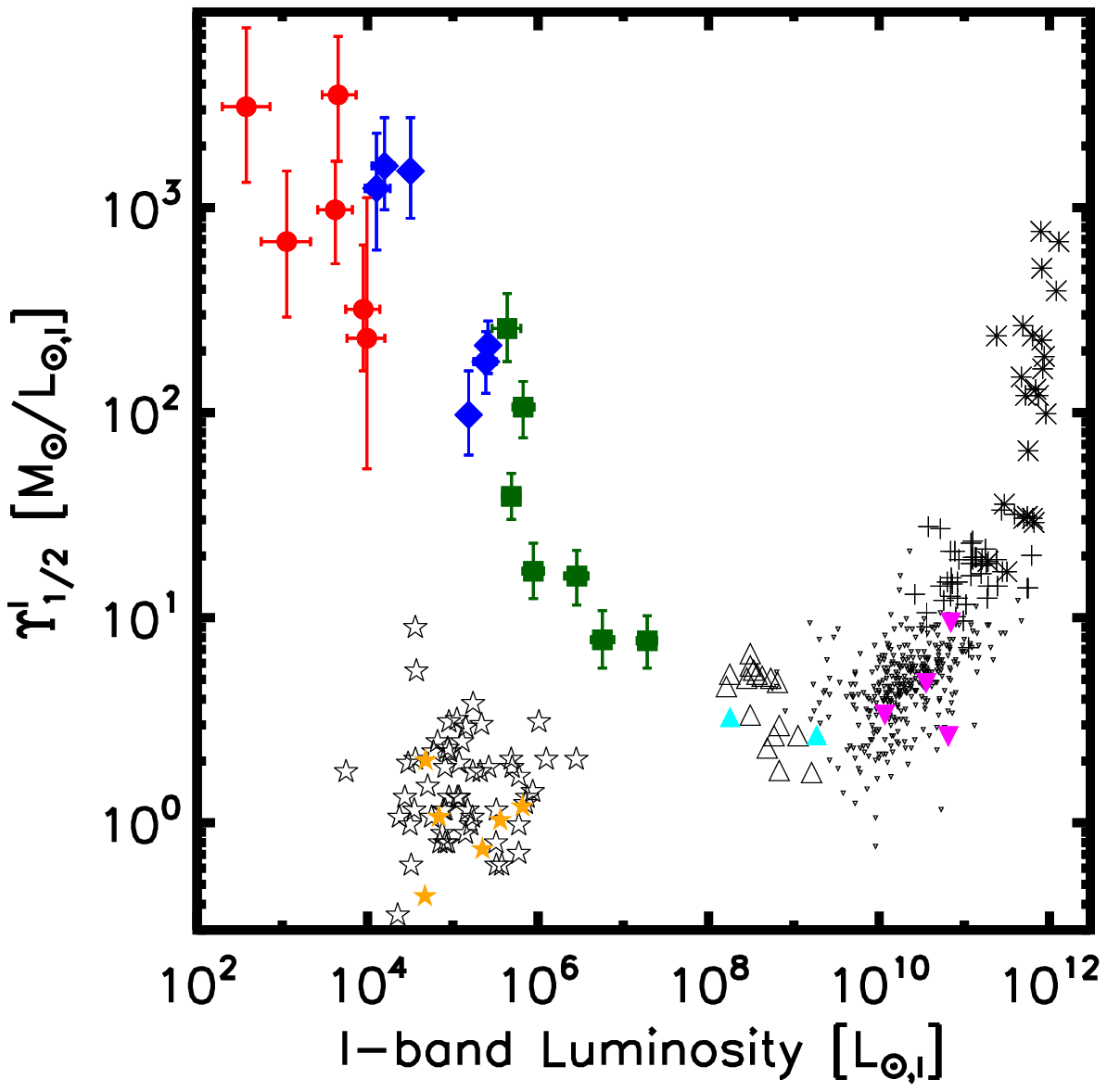}
\caption{See the text for descriptions. The symbols are linked to galaxy types as follows: Milky Way dSphs (squares, diamonds, circles), galactic globular clusters (stars), dwarf ellipticals (triangles), ellipticals (inverted triangles), brightest cluster galaxies (plus signs), and cluster spheroids (asterisks). See Figure 4 of W10 for references.}
\label{fig:manifold}
\end{figure*}

Figure \ref{fig:manifold} examines the relationship between the
half-light mass $\Mhalf$ and the half-light luminosity $\Lhalf = 0.5
\, \LI$ for the full range of dispersion-supported stellar systems in the
universe: globular clusters, dSphs, dwarf ellipticals, ellipticals,
brightest cluster galaxies, and extended cluster spheroids.
\footnote{These findings were expanded upon by \cite{Tollerud_10}
to show that all systems which sit embedded within dark matter halos 
lie along a one dimensional relation within three dimensional space.}

There are several noteworthy aspects to Figure \ref{fig:manifold}, each highlighted in a different way in the three panels. In the middle and right panels the half-light mass-to-light ratios of spheroidal galaxies in the universe demonstrate a minimum at $\moverl \simeq 2-4$ that spans a broad range of luminosities $\LI \simeq 10^{8.5-10.5} \, \Lsun$ and masses $\Mhalf \simeq 10^{9-11} \Msun$. It is interesting to note the offset in the average mass-to-light ratios between L$_\star$ ellipticals and globular clusters, which may suggest that even within $\rhalf$, dark matter may constitute the majority of the mass content of L$_\star$ elliptical galaxies. Nevertheless, it seems that dark matter plays a dominant dynamical role ($\moverl \gtrsim 5$) within $\rhalf$ in only the most extreme systems. The dramatic increase in half-light mass-to-light ratios at both larger and smaller luminosity and mass scales is indicative of a decrease in galaxy formation efficiency in the smallest\footnote{\cite{McGaughWolf_10} recently showed that correlations exist when one explores additional observable properties of dSphs.} and largest dark matter halos. It is worth mentioning that if $\Lambda$CDM is to explain the luminosity function of galaxies a similar trend in the $\Mtwohun$ vs. $L$ relationship must exist. While a different mass variable is presented in Figure \ref{fig:manifold}, the resemblance between the two relationships is striking, and generally encouraging for $\Lambda$CDM.

One may gain some qualitative insight into the physical processes that drive galaxy formation inefficiency in bright vs. faint systems by considering the $\Lhalf$ vs. $\Mhalf$ relation (left panel) in more detail. There exist three distinct power-law regimes $\Mhalf \propto \Lhalf^\aleph$ with $\aleph > 1$, $\aleph \simeq 1$, and $\aleph < 1$ as mass decreases. Over the broad middle range of galaxy masses, $\Mhalf \simeq 10^{9-11} \Msun$, mass and light track each other closely with $\aleph \simeq 1$, while faint galaxies obey $\aleph \simeq 1/2$, and bright elliptical galaxies are better described by $\aleph \simeq 4/3$ transitioning to $\aleph \gg 1$ for the most luminous cluster spheroids. One may interpret the transition from $\aleph > 1$ in bright galaxies to $\aleph < 1$ in faint galaxies as a transition between luminosity-suppressed galaxy formation to mass-suppressed galaxy formation. That is, for faint galaxies ($\aleph < 1$), there does not seem to be a low-luminosity threshold in galaxy formation, but rather behavior closer to a threshold (minimum) mass with variable luminosity. For brighter spheroids with $\aleph > 1$, the increased mass-to-light ratios are driven more by increasing the mass at fixed luminosity, suggestive of a maximum luminosity scale (W10).

Going a step further, Figure \ref{fig:manifold} provides a useful empirical benchmark against which theoretical models can compare. It will be challenging to explain how two of the least luminous MW dSphs have the highest mass-to-light ratios $\moverl \simeq 3,200$ of any the collapsed structures shown, including intra-cluster light spheroids, which reach values of $\moverl \simeq 800$. Not only are the ultra-faint dSphs the most dark matter dominated objects known, given that have even lower baryon-to-dark matter fractions $f_{b} \sim \Omega_{b}/\Omega_{dm} \lesssim 10^{-3}$ than galaxy clusters $f_{b} \simeq 0.1$, W10 points out that ultra-faint dSphs also have higher mass-to-visible light ratios within their stellar extents than even the well-studied galaxy cluster spheroids.
%





\end{document}